\documentclass[letterpaper, 10 pt, conference]{ieeeconf} 
\IEEEoverridecommandlockouts

\usepackage[mode=image]{standalone} 
\usepackage[utf8]{inputenc} 
\usepackage[T1]{fontenc}
\usepackage{url}
\usepackage{ifthen}
\usepackage{cite}
\usepackage[cmex10]{amsmath} 
\usepackage{xspace} 
\usepackage{subcaption}
\usepackage{graphicx,amssymb,amsthm}
\usepackage{xcolor}
\usepackage{booktabs}
\usepackage{bm,color}
\usepackage{tikz,pgfplots,tkz-graph}
\usepackage{balance}

\usepackage{algorithmic}

\usepackage[
	ruled,
	vlined,
	boxed, 
	linesnumbered, 
	commentsnumbered]{algorithm2e}

\SetInd{0.4em}{0.5em}

\usepackage{hyperref}

\theoremstyle{remark} 
\newtheorem{remark}{Remark}

\usetikzlibrary{shapes,arrows}
\usetikzlibrary{calc}

\DeclareMathOperator*{\argmax}{arg\,max}

\DeclareMathOperator*{\PAut}{PAut}
\newcommand{\algorithmsize}{\small}
\newcommand{\define}{\triangleq}
\newcommand{\eps}{\varepsilon}
\newcommand{\transpose}{\intercal}
\newcommand{\dmin}{d_{\mathrm{min}}}
\newcommand{\EbNo}{E_\mathrm{b}/N_0}

\newcommand{\Tmax}{T}

\newcommand{\epsg}{\eps_\mathrm{g}}

\pgfplotsset{compat=newest}%
\usetikzlibrary{positioning,matrix,shapes.multipart,shapes.misc,spy}%
\newcommand{\mat}[1]{\ensuremath{\mathbf{#1}}}
\newcommand{\vect}[1]{\ensuremath{\boldsymbol{#1}}}
\newcommand{\Hstd}{\mat{H}_{\mathrm{std}}}
\newcommand{\Hoc}{\mat{H}_{\mathrm{oc}}}

\tikzset{%
	partial ellipse/.style args={#1:#2:#3}{%
		insert path={+ (#1:#3) arc (#1:#2:#3)}%
	}%
}%
\tikzstyle{standard BP}=[black, mark=diamond*, mark options={solid, fill=white, mark size=2.0pt}, solid]%
\tikzstyle{Nachmani}=[color=blue, mark=*, mark options={solid, fill=white, mark size=1.5pt}, dotted]%
\tikzstyle{annotation}=[fill=white]%

\tikzstyle{diamond marker}=[mark=diamond*, mark options={solid, fill=white, mark size=2.0pt}]
\tikzstyle{triangle marker}=[mark=triangle*, mark options={solid, fill=white, mark size=2.0pt}]
\tikzstyle{square marker}=[mark=square*, mark options={solid, fill=white, mark size=1.3pt}]
\tikzstyle{circle marker}=[mark=*, mark options={solid, fill=white, mark size=1.5pt}]
\tikzstyle{pentagon marker}=[mark=pentagon*, mark options={solid, fill=white, mark size=1.7pt}]
\tikzstyle{star marker}=[mark=asterisk, mark options={solid, fill=white, mark size=1.7pt}]
\tikzstyle{times marker}=[mark=x, mark options={solid, fill=white, mark size=2.0pt}]

\tikzstyle{OSD}=[color=green!50!black, mark=none, solid, thick]%
\tikzstyle{BF Hoc}=[thick, color=black, mark=none, solid]%
\tikzstyle{BF Hstd}=[thick, color=brown!75!black, dashed, pentagon marker]%
\tikzstyle{HD ML}=[thick, color=black, mark=none, dotted]%
\tikzstyle{LBF Hstd}=[ thick, color=red, dashed, square marker]%
\tikzstyle{LBF-NN Hstd}=[thick, color=blue, dashed, triangle marker]%
\tikzstyle{LBF-NN Hoc}=[thick, color=green!50!black, dotted, triangle marker]%
\tikzstyle{WBF Hoc}=[thick, color=magenta, circle marker, dotted]%
\tikzstyle{sort LBF Hstd}=[thick, color=orange, diamond marker, solid]%
\tikzstyle{sort LBF-NN Hstd}=[thick, color=purple, star marker, dashed]%
\tikzstyle{sort LBF-NN Hoc}=[thick, color=red!30!blue, times marker, dotted]%
\tikzstyle{sort LWBF-NN Hstd}=[thick, color=brown!75!black, pentagon marker, dashed]%
\tikzstyle{sort LWBF-NN Hoc}=[thick, color=orange!70!black, square marker, dotted]%

\newif\ifExternalBib 
\ExternalBibfalse

\newif\ifShowNotes 
\ShowNotesfalse

\begin{document}
\title{\LARGE \bf Reinforcement Learning for Channel Coding:\\ Learned Bit-Flipping Decoding} 

\author{%
	Fabrizio Carpi$^{1}$, Christian H\"{a}ger$^{2}$, Marco Martalò$^{3}$, 
	Riccardo Raheli$^{3}$, and Henry D. Pfister$^{4}$
	\thanks{%
		This work was done while F.~Carpi was a student at University of Parma and was visiting Duke University. 
		Preliminary results appeared in the
		thesis \cite{Carpi2018}. 
		The	work of C.~H\"ager was supported by the European Union's Horizon 2020 research and innovation programme under the Marie Sk\l{}odowska-Curie grant No.~749798.
		The work of H.~D.~Pfister was supported in part by the National Science Foundation (NSF) under Grant No.~1718494. 
		Any opinions, findings, conclusions, and recommendations expressed in this material are those of the authors and do not necessarily reflect the views of these sponsors. 
		Please send correspondence to \texttt{henry.pfister@duke.edu}.}
	\thanks{$^{1}$Department of Electrical and Computer Engineering, New York University, Brooklyn, New York, USA}
	\thanks{$^{2}$Department of Electrical Engineering, Chalmers University of Technology, Gothenburg, Sweden}
	\thanks{$^{3}$Department of Engineering and Architecture, University of Parma, Parma, Italy}
	\thanks{$^{4}$Department of Electrical and Computer Engineering, Duke University, Durham, North Carolina, USA}
}

%
%

\maketitle

\begin{abstract}
In this paper, we use reinforcement learning to find effective
decoding strategies for binary linear codes. We start by reviewing
several iterative decoding algorithms that involve a decision-making
process at each step, including bit-flipping (BF) decoding, residual
belief propagation, and anchor decoding. We then illustrate how such
algorithms can be mapped to Markov decision processes allowing for
data-driven learning of optimal decision strategies, rather than
basing decisions on heuristics or intuition. As a case study, we
consider BF decoding for both the binary symmetric and additive white
Gaussian noise channel. Our results show that learned BF decoders can
offer a range of performance--complexity trade-offs for the
considered Reed--Muller and BCH codes, and achieve near-optimal
performance in some cases. We also demonstrate learning convergence
speed-ups when biasing the learning process towards correct decoding
decisions, as opposed to relying only on random explorations and past
knowledge.

\end{abstract}


\section{Introduction}

The decoding of error-correcting codes can be cast as a classification
problem and solved using supervised machine learning. The general idea
is to regard the decoder as a parameterized function (e.g., a neural
network) and learn good parameter configurations with data-driven
optimization \cite{Gruber2017, OShea2017, Nachmani2016, Tallini1995,
Bennatan2018, Kim2018}. Without further restrictions on the code, this only
works well for short codes and typically becomes ineffective for unstructured codes with more than a few hundred codewords.
For linear codes, the problem
simplifies considerably because one has to learn only a single decision region
instead of one region per codeword. One can take advantage of
linearity by using message-passing \cite{Nachmani2016} or syndromes
\cite{Tallini1995, Bennatan2018}.  Still, the problem remains
challenging because good codes typically have complicated decision regions
due to the large number of neighboring codewords. Near-optimal
performance of learned decoders in practical regimes has been
demonstrated, e.g., for convolutional codes \cite{Kim2018}, which
possess even more structure. 

In this paper, we study the decoding of binary linear block codes from
a machine-learning perspective. Rather than learning a direct mapping
from observations to estimated codewords (or bits) in a supervised
fashion, the decoding is done in steps based on individual
bit-flipping (BF) decisions. This allows us to map the problem to a
Markov decision process (MDP) and apply reinforcement learning (RL) to
find good decision strategies. Following \cite{Tallini1995,
Bennatan2018}, our approach is syndrome-based and the state space of
the MDP is formed by all possible binary syndromes, where bit-wise
reliability information can be included for general memoryless
channels. This effectively decouples the decoding problem from the
transmitted codeword.


BF decoding has been studied extensively in the literature and is
covered in many textbooks on modern coding theory, see, e.g.,
\cite{Bossert1986, Kou2001, Zhang2004, Jiang2005, Liu2005, Shan2005},
\cite[Ch.~10.7]{Ryan2009}. Despite its ubiquitous use, and to the best
of our knowledge, the learning approach to BF decoding presented in
this paper is novel. In fact, with the exception of the recent work in
\cite{Wang2019}, we were unable to find references that discuss RL for
channel coding. Thus, we briefly review some other iterative
decoding algorithms, based on sequential decision-making steps, for which RL is applicable. For a comprehensive
survey of RL in the general context of communications, see
\cite{Luong2018}.

\section{Channel Coding Background}
\label{sec:background}





Let $\mathcal{C}$ be an $(N,K)$ binary linear code defined by an $ M
\times N $ parity-check (PC) matrix $\mat{H}$, where $N$ is the code
length, $K$ is the code dimension, and $M \geq N-K$. The code is used
to encode messages into codewords
$\vect{c}=\left(c_1,...,c_N\right)^\transpose$, which are then
transmitted over the additive white Gaussian noise (AWGN) channel
according to $y_{n} = (-1)^{c_{n}} + w_{n}$, where $y_n$ is the $n$-th
component in the received vector
$\vect{y}=\left(y_1,...,y_N\right)^\transpose$, $w_{n}\sim
\mathcal{N}(0,(2 R \EbNo)^{-1})$, $R \define K/N$ is the code rate,
and we refer to $\EbNo$ as the signal-to-noise ratio (SNR). The vector
of hard-decisions is denoted by
$\vect{z}=\left(z_1,...,z_N\right)^\transpose$, i.e., $z_n$ is
obtained by mapping the sign of $y_n$ according to $+1 \to 0$, $-1 \to
1$. If the decoding is based only on the hard-decisions $\vect{z}$,
this scenario is equivalent to transmission over the binary symmetric
channel (BSC). 

%



\subsection{Decision Making in Iterative Decoding Algorithms}

In the following, we briefly review several iterative decoding
algorithms that involve a decision-making process at each step. 

\subsubsection{Bit-Flipping Decoding}


The general idea behind BF decoding is to construct a suitable metric
that allows the decoder to rank the bits based on their reliability
given the code constraints \cite[Ch.~10.7]{Ryan2009}. In its simplest
form, BF uses the hard-decision output $\vect{z}$ and iteratively
looks for the bit that, after flipping it, would maximally reduce the
number of currently violated PC equations. Pseudocode for standard BF
decoding is provided in Alg.~\ref{alg:bit_flipping}, where $\vect{e}_n
\in \mathbb{F}_2^N$ is a standard basis vector whose $n$-th component
is $1$ and all other components are $0$, $\mathbb{F}_2 \define
\{0,1\}$ and $[N] \define \{1,2,\dots, N\}$. BF can be extended to general memoryless channels by
including weights and thresholds to decide which bits to flip at each
step. This is referred to as weighted BF (WBF) decoding, see, e.g.,
\cite{Bossert1986, Kou2001, Zhang2004, Jiang2005, Liu2005, Shan2005},
\cite[Ch.~10.8]{Ryan2009} and references therein.

%
%




\subsubsection{Residual Belief Propagation} 



Belief propagation (BP) is an iterative algorithm where messages are
passed along the edges of the Tanner graph representation of the code.
In general, it is known that sequential message-passing schedules can
lead to faster convergence than standard flooding schedules where
multiple messages are updated in parallel. Residual BP (RBP)
\cite{Elidan2006} is a particular instance of a sequential updating
approach without a predetermined schedule. Instead, the message order
is decided dynamically, where the decisions are based on the
residual---defined as the norm of the difference between the current
message and the message in the previous iteration.  The residual is a
measure of importance or ``expected progress'' associated with sending
the message. In the context of decoding, various extensions of this
idea have been investigated under the name of informed dynamic
scheduling \cite{VilaCasado2010}. 




\subsubsection{Anchor Decoding} 

Consider the iterative decoding of product codes\footnote{Given a
linear code $\mathcal{C}$ of length $n$, the product code of
$\mathcal{C}$ is the set of all $n \times n$ arrays such that each row
and column is a codeword in $\mathcal{C}$.} over the BSC, where the
component codes are iteratively decoded in some fixed order. For this
algorithm, undetected errors in the component codes, so-called
miscorrections, significantly affect the performance by introducing
additional errors into the iterative decoding process. To address this
problem, anchor decoding (AD) was recently proposed in
\cite{Haeger2018tcom}. The AD algorithm exploits conflicts due to
miscorrections where two component codes disagree on the value of a
bit. After each component decoding, a decision is made
based on the number of conflicts whether the decoding outcome is
indeed reliable. This can lead to backtracking previous component decoding outcomes and to the designation
of reliable component codes as anchors. 


%

\subsection{Decision Making Through Data-Driven Learning}

While the above decoding algorithms appear in seemingly different
contexts, the sequential decision-making strategies in the underlying
iterative processes are quite similar. Decisions are typically made in
a greedy fashion based on some heuristic metric that assesses the
quality of each possible action. As concrete examples for this metric,
we have 

\begin{itemize}
	\item the decrease in the number of violated PC equations in BF
		decoding, measuring the reliability of bits;

	\item the residual in RBP, measuring expected progress and the
		importance of sending messages;

	\item the number of conflicts in AD, measuring the
		likelihood of being miscorrected. 

\end{itemize}

In the next section, we review MDPs which provide a mathematical
framework for modeling decision-making in deterministic or random
environments. MDPs can be used to obtain optimal decision-making
strategies, effectively replacing heuristics with data-driven learning
of optimal metrics. 

\newcommand{\rightcomment}[1]{\tcp*[r]{#1}}
\newcommand\mycommfont[1]{\scriptsize\ttfamily\textcolor{blue}{#1}}
\SetCommentSty{mycommfont}

\setlength{\textfloatsep}{10pt}
\begin{algorithm}[t]
	\algorithmsize
	\DontPrintSemicolon
	\SetKw{ShortFor}{for}
	\SetKw{KwBreak}{break}
	\SetKw{MyWhile}{while}
	\SetKw{MyIf}{if}
	\SetKw{MySet}{set}
	\SetKw{MyElse}{else}
	\SetKw{MyCompute}{compute}
	\SetKw{KwEach}{each}
	\SetKw{KwAnd}{and}
	
	\KwIn{hard decisions $\vect{z}$, parity-check matrix $\mat{H}$}
	\KwOut{estimated codeword $\hat{\vect{c}}$}
	$\hat{\vect{c}} \leftarrow \vect{z}$\;
	\While{$\mat{H} \hat{\vect{c}} \neq \vect{0} $ \KwAnd \emph{max.~iterations not exceeded}}{
		$V \leftarrow \sum_{m=1}^{M} s_m$, where $\vect{s} = \mat{H}
		\hat{\vect{c}}$ \rightcomment{no.~unsat~checks}
		\For{$n = 1, 2, \dots, N$}{
			$Q_n \leftarrow V - \sum_{m=1}^{M} s_m$, where $\vect{s} =
			\mat{H} (\hat{\vect{c}} + \vect{e}_n)$ \;
		}
		update $\hat{\vect{c}} \leftarrow \hat{\vect{c}} + \vect{e}_n $, where $n = \arg\max_{n \in
		[N]} Q_n$\;
	}
	\caption{ Bit-Flipping Decoding }
	\label{alg:bit_flipping}
\end{algorithm}

\section{Markov Decision Processes}

A time-invariant MDP is a Markov random process $S_0$, $S_1$, $\dots$
whose state transition probability $P(s' | s, a) \define
\mathbb{P}(S_{t+1} = s' | S_t = s , A_t = a)$ is affected by the
action $A_t$ taken by an agent based only on knowledge of past events.
Here, $s, s' \in \mathcal{S}$ and $a \in \mathcal{A}$, where
$\mathcal{S}$ and $\mathcal{A}$ are finite sets containing all
possible states and actions. The agent also receives a reward $R_t =
R(S_t, A_t, S_{t+1})$ which depends only on the states $S_t$,
$S_{t+1}$ and the action $A_t$. The agent's decision-making process is
formally described by a policy $\pi : \mathcal{S} \to \mathcal{A}$,
mapping observed states to actions. The goal is to find an optimal
policy $\pi^*$ that returns the best action for each possible state in
terms of the total expected discounted reward
$\mathbb{E}\left[\sum_{t=0}^\infty \gamma^t R_t\right]$, where $0 <
\gamma < 1$ is the discount factor for future rewards. 


If the transition and reward probabilities are known, dynamic
programming can be used to compute optimal policies. If this is not
the case, optimal policies can still be discovered through repeated
interactions with the environment, assuming that the states and
rewards are observable. This is known as RL.  In the following, we
describe two RL algorithms which will be used in the next sections. 

\subsection{Q-learning}

The most straightforward instance of RL is called Q-learning
\cite{Watkins1989}, where the optimal policy is defined in terms of
the Q-function $Q : \mathcal{S} \times \mathcal{A} \to \mathbb{R}$
according to 
\begin{align}
	\pi^*(s) = \argmax_{a \in \mathcal{A}} Q(s,a). 
\end{align}
The Q-function measures the quality of actions and is formally defined
as the expected discounted future reward when being in state $s$,
taking action $a$, and then acting optimally. The key advantage of the
Q-function is that it can be iteratively estimated from observations
of any ``sufficiently-random'' agent.  Pseudocode for Q-learning is
given in Alg.~\ref{alg:q_learning}, where a popular choice for
generating the actions in line 5 is 
\begin{equation}
	\label{eq:eps-greedy}
	a =	\begin{cases}
		\text{unif.~random over $\mathcal{A}$} &\quad \text{w.p.~}
		\eps\\
			\argmax_{a} Q(s,a) &\quad \text{w.p.~} 1- \eps.
		\end{cases}
\end{equation}
This is referred to as $\eps$-greedy exploration. For any $0 < \eps <
1$, this strategy is sufficient to allow Q-learning to eventually
explore the entire state/action space. In the next section, we also
describe an alternative exploration strategy for our application that
can converge faster than $\eps$-greedy exploration. 

To motivate the update equation in line 7 of
Alg.~\ref{alg:q_learning}, we note that the Q-function can be
recursively expressed as 
\begin{align}
	\!\!Q(s,a) = \sum_{s'} P(s'|s,a) \left( \! R(s,a,s') + \gamma \max_{a' \in
	\mathcal{A}} Q(s', a') \! \right) \! . 
\end{align}
This expression forms the theoretical basis for Q-learning which
converges to the true Q-function under certain conditions\footnote{For example, if $R(s,a,s')$ depends non-trivially on $s'$, then $\alpha$ must decay to zero at sufficiently slow rate.}. For a more
details, we refer the reader to \cite{Watkins1989, Sutton1998}. 

\begin{algorithm}[t]
	\algorithmsize
	\DontPrintSemicolon
	\SetKw{ShortFor}{for}
	\SetKw{KwBreak}{break}
	\SetKw{MyWhile}{while}
	\SetKw{MyIf}{if}
	\SetKw{MySet}{set}
	\SetKw{MyElse}{else}
	\SetKw{MyCompute}{compute}
	\SetKw{KwEach}{each}
	\SetKw{KwAnd}{and}
	
	\KwIn{learning rate $\alpha$, discount factor $\gamma$}
	\KwOut{estimated Q-function}
	initialize $Q(s,a) \leftarrow 0$ for all $s \in \mathcal{S}$, $a \in \mathcal{A}$\;
	\For{$i = 1, 2, \dots$}{
	initialize starting state $s$ \rightcomment{restart the MDP}
	\While{$s$ \emph{is not terminal}}{
		choose action $a$ \rightcomment{$\eps$-greedy
		\eqref{eq:eps-greedy} or
		$(\eps, \epsg)$-goal \eqref{eq:eps-goal}}
		execute $a$, observe reward $r$ and next state $s'$\;
		\hspace{-0.25em}$Q(s,a) \leftarrow (1-\alpha)Q(s,a) + \alpha(r + \gamma \max_{a'
		\in \mathcal{A}} Q(s',a'))$\;
		$s \leftarrow s'$\;
	}
	}
	\caption{ Q-learning }
	\label{alg:q_learning}
\end{algorithm}

\subsection{Fitted Q-learning with Function Approximators}


For standard Q-learning, one must store a table of $|\mathcal{S}|
\times |\mathcal{A}|$ real values. This will be infeasible if either
set is prohibitively large. The idea of fitted Q-learning is to learn
a low-complexity approximation of $Q(s,a)$ \cite{Sutton1998}. Let
$Q_\theta(s,a)$ be an approximation of the Q-function, parameterized
by $\theta$. Fitted Q-learning alternates between simulating the MDP
and updating the current parameters to obtain a better estimate of the
Q-function. In particular, assume that we have simulated and stored
$B$ transition tuples $(s, a, r, s')$ in a set $\mathcal{D}$. Then,
updating the parameters $\theta$ is based on reducing the empirical
loss 
\begin{align}
	\label{eq:loss}
	\!\!\!\mathcal{L}_{\mathcal{D}}(\theta) =\!\! \sum_{(s,a,r,s') \in
	\mathcal{D}} \left(r + \gamma \max_{a' \in \mathcal{A}} Q_\theta(s', a') -
	Q_\theta(s,a) \right)^2 \! \! . 
\end{align}
Pseudocode for fitted Q-learning is provided in
Alg.~\ref{alg:fitted_q_learning}, where gradient descent is used to
update the parameters $\theta$ based on the loss \eqref{eq:loss}. It
is now common to choose $Q_\theta(s,a)$ to be a (deep) neural network
(NN), in which case $\theta$ are the network weights and fitted
Q-learning is called deep Q-learning. 

\begin{algorithm}[t]
	\algorithmsize
	\DontPrintSemicolon
	\SetKw{ShortFor}{for}
	\SetKw{KwBreak}{break}
	\SetKw{MyWhile}{while}
	\SetKw{MyIf}{if}
	\SetKw{MySet}{set}
	\SetKw{MyElse}{else}
	\SetKw{MyCompute}{compute}
	\SetKw{KwEach}{each}
	\SetKw{KwAnd}{and}
	
	\KwIn{learning rate $\alpha$, batch size $B$}
	\KwOut{parameterized estimate of the Q-function}
	initialize parameters $\theta$ and $\mathcal{D} \leftarrow
	\emptyset $ \;
	\For{$i = 1, 2, \dots$}{
	initialize starting state $s$ \rightcomment{restart the MDP}
	\While{$s$ \emph{is not terminal}}{
		choose action $a$ \rightcomment{$\eps$-greedy
		\eqref{eq:eps-greedy} or
		$(\eps, \epsg)$-goal \eqref{eq:eps-goal}}
		execute $a$, observe reward $r$ and next state $s'$\;
		store transition $(s, a, r, s')$ in $\mathcal{D}$\;
		$s \leftarrow s'$\;
		\If{ $|\mathcal{D}| = B$}{
		$\theta \leftarrow \theta - \alpha \nabla_\theta
		\mathcal{L}_{\mathcal{D}}(\theta) $ \rightcomment{see
		\eqref{eq:loss} for def.~of $\mathcal{L}_\mathcal{D}$}
		empty $\mathcal{D}$\;
	}%
	}%
	}%
	\caption{ Fitted Q-learning }
	\label{alg:fitted_q_learning}
\end{algorithm}


\section{Case Study: Bit-Flipping Decoding}
\label{sec:case_study}

In this section, we describe how BF decoding can be mapped to an MDP.
In general, this mapping involves multiple design choices that
affect the results. We therefore also comment on alternative
choices and highlight some potential pitfalls that we encountered
during this process.  

\subsection{Theoretical Background}

We start by reviewing the standard maximum-likelihood (ML) decoding
problem for a binary linear code $\mathcal{C} \subseteq
\mathbb{F}_2^N$ over general discrete memoryless channels. The resulting
optimization problem forms the basis for the reward function that is
used in the MDP. To that end, consider a collection of $N$ discrete
memoryless channels described by conditional probability density
functions $\{ P_{Y_n|C_n}(y_n|c_n) \}_{n \in [N]}$, where $c_n \in
\mathbb{F}_2$ is the $n$-th code bit and $y_n$ is the $n$-th channel
observation. The ML decoding problem can be written as 
\begin{align}
	\!\!\! \argmax_{\vect{c} \in \mathcal{C}} \prod_{n=1}^N \! P_{Y_n|C_n} (y_n | c_n)
	=
	\argmax_{\vect{c} \in \mathcal{C}} \sum_{n=1}^N (-1)^{c_n} \lambda_n,\! 
\end{align}
where 
\begin{align}
	\lambda_n \define \ln \frac{ P_{Y_n|C_n}(y_n|0) }{
	P_{Y_n|C_n}(y_n|1) }
\end{align}
is the channel log-likelihood ratio (LLR). Equivalently, one can
rewrite the maximization over all possible codewords in terms of error
patterns as
\begin{align}
	&\argmax_{\vect{e} \,:\, \vect{z} + \vect{e} \in \mathcal{C}} \sum_{n=1}^N
	(-1)^{z_n}(-1)^{e_n} \lambda_n \\
	&=\argmax_{\vect{e} \,:\, \vect{z} + \vect{e} \in \mathcal{C}} \sum_{n=1}^N
	(-1)^{e_n} |\lambda_n| \\
	&=\argmax_{\vect{e} \,:\, \mat{H}\vect{e} = \vect{s}} \sum_{n=1}^N
	(-1)^{e_n} |\lambda_n|\\
	&=\argmax_{\vect{e} \,:\, \mat{H}\vect{e} = \vect{s}} \sum_{n=1}^N
	-{e_n} |\lambda_n| 
\end{align}
where $\vect{s}=\mat{H}\vect{z}$ is the observed syndrome.

Now, consider a multi-stage process where bit $a_t$ is flipped during
the $t$-th stage until the syndrome of the bit-flip pattern matches $\vect{s}$. 
In this case, the optimization becomes
\begin{align}
	\label{eq:ml}
	\argmax_{\tau, a_1, \dots, a_\tau \,:\, \sum_{t=1}^\tau \vect{h}_{a_t} =
	\vect{s}} \sum_{t = 1}^\tau - |\lambda_{a_t}|,
\end{align}
where $\vect{h}_{n}$ is the $n$-th column of the parity-check matrix
$\mat{H}$. By interpreting $-|\lambda_{a_t}|$ as a reward, one can see
that the objective function in \eqref{eq:ml} has the same form as the
cumulative reward (without discount) in an MDP. The following points
are worth mentioning: 
\begin{itemize}
	\item For the BSC, all LLRs have the same magnitude and
		\eqref{eq:ml} returns the shortest flip pattern that matches the
		observed syndrome.

	\item For general channels, \eqref{eq:ml} returns the shortest
		\emph{weighted} flip pattern that matches the syndrome, where
		the weighting is done according to the channel LLRs.  In other
		words, the incurred penality for flipping bit $a_t$ is directly
		proportional to the reliability of the corresponding received
		bit. 

	\item If a bit is flipped multiple times, then there must be a
		shorter bit-flip sequence with lower cost and the same syndrome.
		Therefore, it is sufficient to only consider  flip patterns that contain distinct
		bits. 

\end{itemize}




\subsection{Modeling the Markov Decision Process}

\subsubsection{Choosing Action and State Spaces}

We assume that the action $A_t$ encodes which bit is flipped in the
received word at time $t$. Since there are $N$ possible choices, we
simply use $\mathcal{A} = \{1,2,\dots, N\} \define [N]$.  The state
space $\mathcal{S}$ is formed by all possible binary syndromes of
length $M$. The initial state $S_0$ is the syndrome $\mat{H}\vect{z}$
and the next state is formed by adding the $A_t$-th column of
$\mat{H}$ to the current state. The transition probabilities
$P(s'|s,a)$ therefore take values in $\{0,1\}$, i.e., the MDP is
deterministic. The all-zero syndrome corresponds to a terminal state.
We also enforce a limit of at most $T$ bit-flips per codeword. After
this, we exit the current iteration and a new codeword will be
decoded.\footnote{Strictly speaking, the resulting process is not an
MDP unless the time $t$ is included in the state space.}


\begin{remark}
	For the BSC, we also tried (unsuccessfully) to learn BF decoding with fitted
	Q-learning directly from the channel observations using the state space $\mathbb{F}_2^N$. 
\end{remark}

\begin{remark}
	For the AWGN channel, the state space can be extended by including
	the reliability vector $\vect{r} = | \vect{y} |$, similar to the
	setup in \cite{Bennatan2018}. In this case, each state would
	correspond to a tuple $(\vect{s}, \vect{r})$, where
	$\vect{s}\in\mathbb{F}_2^M$ and $\vect{r}$ remains constant during
	decoding. In this paper, we follow a different strategy for BF
	decoding over the AWGN channel which relies on permuting the bit
	positions based on their reliability and subsequently discarding
	the channel LLRs prior to decoding. This approach is described in
	Sec.~\ref{sec:sorted} and does not require any modifications to the
	state space. 
\end{remark}

\subsubsection{Choosing the Reward Strategy}


A natural reward function for decoding is to return $1$ if the
codeword is decoded correctly and $0$ otherwise. This would imply that
an optimal policy minimizes the codeword error rate. However, the
reward is only allowed to depend on the current/next state and the
action, whereas the transmitted codeword and its estimate are defined
outside the context of the MDP. Based on \eqref{eq:ml} and the
discussion in the previous subsection, we instead use the reward
function
\begin{align}
	\label{eq:reward_function}
	\!\!R(s,a,s') = \begin{cases}
		-c |\lambda_a| +1 &\! \text{if $s' = \vect{0}$}\\ 
		-c |\lambda_a|  &\! \text{otherwise },
	\end{cases}
\end{align}
where $c > 0$ is a scaling factor. The additional reward for matching
the syndrome is required to prevent the decoder from just flipping the
bits where $|\lambda_a|$ is minimal. For example, it could happen that
a single error in position $a$ with large $|\lambda_a|$ matches the
syndrome, but instead one chooses to flip $\Tmax$ bits with small
absolute LLRs. The scaling factor $c$ is chosen such that the
syndrome-matching reward $+1$ always dominates the expected
cummulative term $-\sum_{t=1}^T{c |\lambda_{a_t}|}$. As an example,
for the BSC, $c$ is chosen such that the reward function becomes
\begin{align}
	\!\!R(s,a,s') = \begin{cases}
		-\frac{1}{T}+1 &\! \text{if $s' = \vect{0}$}\\ 
		-\frac{1}{T} &\! \text{otherwise}.
	\end{cases}
\end{align}
This reward function allows us to interpret optimal BF decoding as a
``maze-playing game'' in the syndrome domain where the goal is to find
the shortest path to the all-zero syndrome. Applying a small negative
penalty for each step is a standard technique to encourage short
paths. Another alternative in this case is to choose a small discount
factor $\gamma < 1$. 


\subsubsection{Choosing the Exploration Strategy}


Compared to \eqref{eq:eps-greedy}, we propose another exploration
strategy as follows. Let $\vect{e}$ be the current error
pattern, i.e., the channel error pattern plus any bit-flips that have
been applied so far. Then, with probability $\epsg$, we choose the
action randomly from $\operatorname{supp}(\vect{e}) \define \{i
\in [N] \,|\, e_i = 1\}$, i.e., we flip one of the incorrect
bits. When combined with $\eps$-greedy exploration, we refer to this
as $(\eps, \epsg)$-goal exploration, where $\eps, \epsg > 0$ and $0 <
\eps + \epsg < 1$:
\begin{equation}
	\label{eq:eps-goal}
	\!\! a =	\begin{cases}
		\text{unif.~random over $\mathcal{A}$} & \text{w.p.~}
		\eps\\
			\text{unif.~random over
			$\operatorname{supp}(\vect{e})$ } & \text{w.p.~} \epsg\\
			\arg \max_{a} Q(s,a) & \text{w.p.~} 1 - \eps  -  \epsg.
		\end{cases}
\end{equation}

\begin{remark}
	It may seem that biasing actions towards flipping erroneous bits
	leads to a form of supervised learning where the learned decisions
	merely imitate ground-truth decisions. To see that this is not
	exactly true, consider transmission over the BSC where the error
	pattern has weight $\dmin - 1$ (where $\dmin$ is the minimum distance of the code) and the observation is at distance
	$1$ from a codeword $\tilde{\vect{c}}$. Then, the optimal decision
	is to flip the bit that leads to $\tilde{\vect{c}}$, whereas
	flipping an erroneous bit is suboptimal in terms of expected future
	reward, even though it moves us closer to the transmitted codeword
	$\vect{c} \neq \tilde{\vect{c}}$. 
\end{remark}


\subsubsection{Choosing the Function Approximator}



We use fully-connected NNs with one hidden layer to represent
$Q_\theta(s,a)$ in fitted Q-learning. In particular, the NN
$\vect{f}_\theta$ maps syndromes to length-$N$ vectors
$\vect{f}_\theta(\vect{s}) \in \mathbb{R}^N$ and the Q-function is
given by $Q_\theta(s,a) = [\vect{f}_\theta(\vect{s})]_a$, where
$[\cdot]_n$ returns the $n$-th component of a vector and $\vect{s}$ is
the syndrome for state $s$. The NN parameters are summarized in
Tab.~\ref{tab:network_parameters}. In future work, we plan to
explore other network architectures, e.g., multi-layer NNs or graph
NNs based on the code's Tanner graph. 




\begin{table}
\centering
\caption{Neural network parameters}
\begin{tabular}{c|ccc}
\toprule
layer               & input & hidden  & output  \\ \midrule
number of neurons   & $M$ &  $500$ / $1500$  & $N$  \\ 
activation function & - & ReLU & linear    \\ 
\bottomrule
\end{tabular}
\label{tab:network_parameters}
\end{table}

%


\section{Learned Bit-Flipping with Code Automorphisms}
\label{sec:sorted}

Let $\mathcal{S}_N$ be the symmetric group on $N$ elements so that
$\pi \in \mathcal{S}_N$ is a bijective mapping (or permutation) from
$[N]$ to itself.\footnote{For a group $(G, \circ)$, we also informally
refer to the set $G$ as the group. In our context, the group operation
$\circ$ represents function composition defined by $(\pi \circ \sigma)(i)= \pi(\sigma(i))$.} The permutation
automorphism group of a code $\mathcal{C}$ is defined as
$\PAut(\mathcal{C}) \define \{\pi \in \mathcal{S}_N \,|\, \vect{x}^\pi
\in \mathcal{C}, \forall \vect{x} \in \mathcal{C}\}$, where
$\vect{x}^\pi$ denotes a permuted vector, i.e., $x_i^\pi =
x_{\pi(i)}$. The permutation automorphism group can be exploited in
various ways to improve the performance of practical decoding
algorithms, see, e.g., \cite{Jiang2004}, \cite{Halford2006}.  In the
context of learned decoders, the authors in \cite{Bennatan2018}
propose to permute the bit positions prior to decoding (and unpermute
after) such that the channel reliabilities are approximately sorted.
If the applied permutations are from $\PAut(\mathcal{C})$, the decoder
simply decodes a permuted codeword, rather than the transmitted one.
The advantage is that certain bit positions are now more reliable than
others due to the (approximate) sorting. This can be advantageous in
terms of optimizing parameterized decoders because of the additional
structure that the decoder can rely on \cite{Bennatan2018}.  

\subsection{A Permutation Strategy for Reed--Muller Codes}

In \cite{Bennatan2018}, the permutation preprocessing approach is
applied for Bose--Chaudhuri--Hocquenghem (BCH) codes and permutations
are selected from $\PAut(\mathcal{C})$ such that the total
reliabilities of the first $K$ permuted bit positions are maximized,
see \cite[App.~II]{Bennatan2018} for details. In the following, we
propose a variation of this idea for RM codes. In particular, our goal
is to find a permutation that sends as many as possible of the
\emph{least reliable} bits to positions $\{0, 1, 2, 4, \dots,
2^{m-1}\} \define \mathcal{B}$. Recall that the automorphism group of
RM$(r,m)$ is the general affine group of order $m$ over the binary
field, denoted by AGL$(m,2)$ \cite[Th.~24]{MacWilliams1977}.  The
group AGL$(m,2)$ is the set of all operators of the form
\begin{align}
	\label{eq:affine}
	T(\vect{v}) = \mat{A} \vect{v} + \vect{b},
\end{align}
where $\mat{A} \in \mathbb{F}_2^{m \times m}$ is an invertible binary
matrix and $\vect{b}, \vect{v} \in \mathbb{F}_2^m$. By interpreting
the vector $\vect{v}$ as the binary representation of a bit position
index, \eqref{eq:affine} defines a permutation on the index set $\{0,
1, \dots, N-1\}$ and thus on $[N]$.

A set of vectors $\{\bm{v}_0,\bm{v}_1,\ldots,\bm{v}_{m}\}$ is called
\emph{affinely independent} if and only if the set
$\{\bm{v}_1-\bm{v}_0,\ldots,\bm{v}_{m}-\bm{v}_0\}$ is linearly
independent. The binary representations of the indices in
$\mathcal{B}$ correspond to the all-zero vector and all unit vectors
of length $m$. One can verify that they are affinely independent. The
proposed strategy relies on the fact that, for any given set of $m+1$
affinely independent bit positions (in the sense that their binary
representation vectors are affinely independent), there always exists
a permutation in AGL$(m,2)$ such that the bit positions are mapped to
$\mathcal{B}$ in any desired order. In particular, we perform the
following steps to select the permutation prior to decoding:
\begin{enumerate}
	\item Let $\pi$ be the permutation that sorts the reliability
		vector $\vect{r} = |\vect{y}|$, i.e., $\vect{r}^\pi$ satisfies
		$r_i^\pi < r_j^\pi$ $\iff$ $i < j$.  

	\item Find the first $m+1$ affinely independent indices for $\pi$
		(e.g., using Gaussian elimination) and denote their binary
		representations by $\vect{v}_0, \vect{v}_1, \dots, \vect{v}_m$. 

	\item The permutation is then defined by \eqref{eq:affine}, where
		$\vect{b} = \vect{v}_0$ and the columns of $\mat{A}$ are 
		$\vect{v}_1 - \vect{v}_0, \dots, \vect{v}_m - \vect{v}_0$. 

\end{enumerate}

\subsection{(Approximate) Sort and Discard}

For the learned BF decoders over the AWGN channel, our approach is to
first apply the permutation strategy described in the previous section
and subsequently discard the channel LLRs. From the perspective of the
decoder, this scenario can be modeled as $N$ parallel BSCs, where the
crossover probabilities for the bit positions in $\mathcal{B}$ satisfy
$p_0 > p_1 > p_2 > p_4 > \dots > p_{2^{m-1}}$. This is related to
approaches where channel reliabilities are used to mark highly
reliable and/or unreliable bit positions, while the actual decoding is
performed without knowledge of the reliability values using
hard-decision decoding, see, e.g., \cite{Lei2018}. 



\begin{figure}[t]
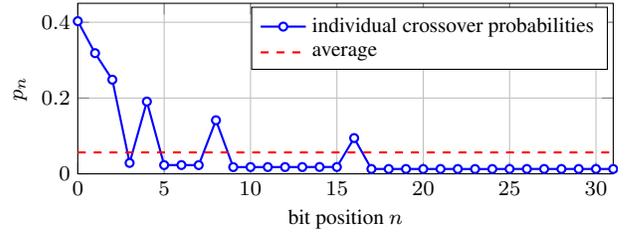

	\centering 
	\includestandalone{pe} 
	\caption{BSC crossover probabilities after the proposed permutation
	strategy for RM$(32,16)$ at $\EbNo =
	4$\,dB. }
	\label{fig:pe}
\end{figure}

The absolute values of the channel LLRs for the parallel BSCs used in
the reward function \eqref{eq:reward_function} are given by
\begin{align}
	|\lambda_n| = \log \frac{1 - p_n}{p_n}, 
\end{align}
where $p_n$ is the crossover probability of the $n$-th BSC. The
individual crossover probabilities can be determined via Monte Carlo
estimation before the RL starts. For example, Fig.~\ref{fig:pe} show
the expected crossover probabilities after applying the proposed
permutation strategy for RM$(32,16)$ assuming transmission at $\EbNo =
4$\,dB. 

%

\begin{figure}[t]
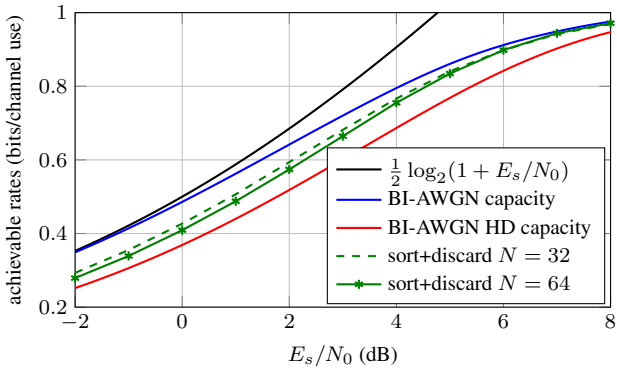

	\centering 
	\includestandalone{capacity} 
	\caption{Estimation of
		achievable information rates when applying the proposed
		permutation strategy for RM codes and subsequently discarding the reliability values.
		(BI-AWGN: binary-input AWGN, HD: hard decision)}
	\label{fig:capacity}
	\vspace{-0.3em}
\end{figure}

\begin{remark}
One can estimate the capacity of strategies that permute the received
bits using the reliabilities and then discard them.
Fig.~\ref{fig:capacity} shows the estimated information rates for the
proposed strategy obtained
via Monte Carlo averaging. Our results show that a significant fraction of the achievable information rate is preserved,
especially for high-rate codes. For permutations restricted to
AGL$(m,2)$, this is less effective as the blocklength
increases because the fraction of sorted channels satisfies $(m+1)/N =
(\log_2(N)+1)/N$. 
\end{remark}




\vspace{1mm}
\section{Results}
\label{sec:results}

In this section, numerical results are presented for learned BF
(LBF) decoders\footnote{$\mat{H}$-matrices and source code for the simulations are
available online at 
\url{https://github.com/fabriziocarpi/RLdecoding}.
We first used our own Tensorflow RL implementation and later switched to \texttt{RLlib} \cite{Liang2018rllib} in order to use multi-core parallelism for training rollouts.} for the following RM and BCH codes: 

\begin{itemize}
	\item RM$(32,16)$ with the standard $16 \times
			32$ PC matrix $\Hstd$ and overcomplete $620 \times 32$ PC matrix $\Hoc$
		whose rows are all minimum-weight dual codewords, see
		\cite{Bossert1986, Santi2018}


	\item RM$(64, 42)$ with the standard $22 \times 64$ PC matrix $\Hstd$ and overcomplete $2604 \times 64$ PC matrix $\Hoc$

	\item BCH$(63, 45)$ with the standard $18
			\times 63$ circulant PC matrix $\Hstd$ and overcomplete $189 \times 63$ PC matrix $\Hoc$




%


	\item RM$(128,99)$ with the standard $29 \times
			128$ PC matrix $\Hstd$ and overcomplete $10668 \times 128$ PC matrix $\Hoc$
		

\end{itemize}
For some of the considered codes, standard table Q-learning is
feasible.
For example, RM$(32,16)$ has $|\mathcal{S}| = 2^{16} =
65536$ and $|\mathcal{A}| = 32$ so the Q-table has
$|\mathcal{S}||\mathcal{A}| \approx 2 \cdot 10^6$ entries. 



\subsection{Training Hyperparameters}

In the following, we set the maximum number of decoding iterations to
$T = 10$ and the discount factor to $\gamma = 0.99$. For standard table
Q-learning, the $(\eps, \epsg)$-goal exploration strategy is adopted
with fixed $\eps = 0.6$, $\epsg = 0.3$, and learning rate $\alpha =
0.1$. For fitted Q-learning based on NNs, we use $\eps$-greedy
exploration where $\eps$ is linearly decreased from $0.9$ to $0$ over
the course of $0.9 K$ learning episodes (i.e., number of decoded
codewords), where the total number of episodes $K$ depends on the
scenario. For the gradient optimization, the Adam optimizer is used
with a batch size of $B=100$ and learning rate $\alpha = 3 \cdot
10^{-5}$. The training SNR for both standard Q-learning and fitted
Q-learning is fixed at $\EbNo = 5\,$dB for RM$(128,99)$ and $\EbNo =
4\,$dB for all other codes. In general, better performance may be
obtained by re-optimizing parameters for each SNR or by adopting
parameter adapter networks that dynamically adapt the network
parameters to the SNR \cite{Lian2019isit}. 




\subsection{Learning Convergence in Q-Learning}

\begin{figure}[t]
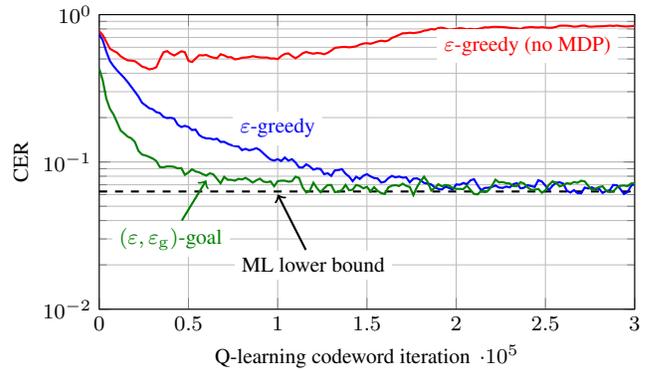

	\centering
	\includestandalone{learning_curve}
	\caption{Q-learning convergence for RM$(32,16)$ on the BSC
	(crossover prob.~$0.0565$ corresponding to $\EbNo = 4\,$dB)
	assuming $T = 10$, $\alpha = 0.1$,
	$\gamma = 1.0$, and $\eps = 0.9$ for $\eps$-greedy and $\eps =
	0.6$, $\epsg = 0.3$ for $(\eps, \epsg)$-goal. }
	\label{fig:learning_curve}
	\vspace{-1mm}
\end{figure}

We start by comparing the learning convergence of the proposed
exploration strategy \eqref{eq:eps-goal} to the $\eps$-greedy
exploration for standard Q-learning assuming RM$(32,16)$ over the BSC.
In Fig.~\ref{fig:learning_curve}, the obtained performance in terms of
codeword error rate (CER) is shown as a function of the Q-learning
iteration. The shown learning curves are generated as follows. During
Q-learning, we always decode first the new channel observations (line
3 of Alg.~\ref{alg:q_learning}) with the current Q-function without
exploration and save the binary outcome (success/failure). Then, we
plot a moving average (window size $5000$) of the outcomes to
approximate the CER. It can be seen that the proposed strategy
converges significantly faster than $\eps$-greedy exploration. We also
show a learning curve for training when a reward of $1$ is given only for finding the transmitted codeword; in this case, however, the process is not an MDP (see
Sec.~\ref{sec:case_study}) and the performance can become worse
during training.  

\subsection{Binary Symmetric Channel}

\begin{figure*}[t]
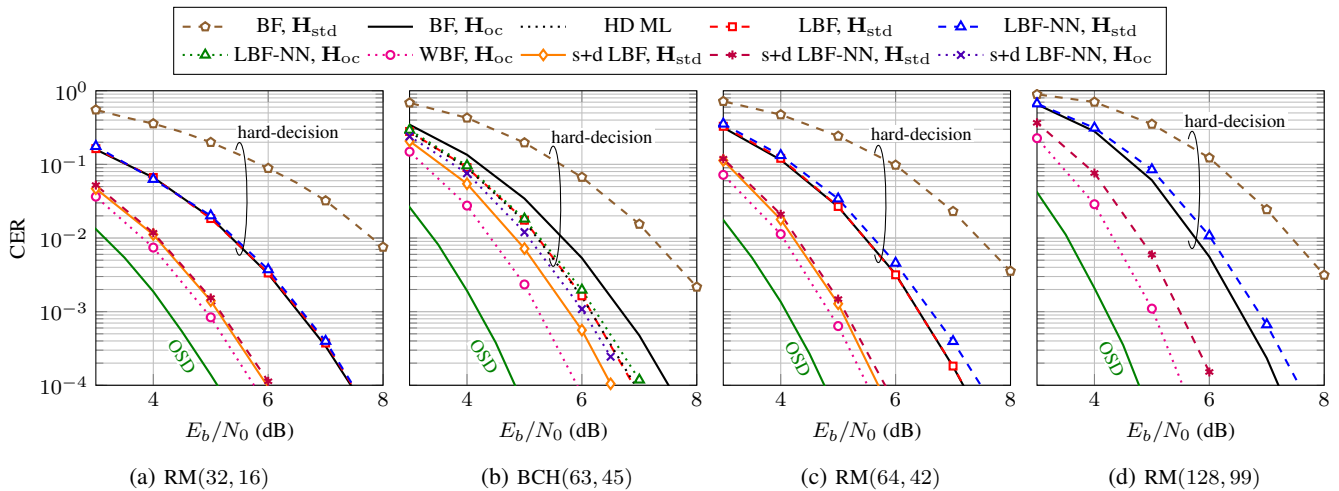

	\centering%
	\includestandalone{legend_CER}\\
	\subcaptionbox{\footnotesize RM$(32, 16)$}{\includestandalone{RM_2_5_CER}}
	\subcaptionbox{\footnotesize BCH$(63, 45)$}{\includestandalone{BCH_63_45_CER}}
	\subcaptionbox{\footnotesize RM$(64, 42)$}{\includestandalone{RM_3_6_CER}}
	\subcaptionbox{\footnotesize RM$(128, 99)$}{\includestandalone{RM_4_7_CER}}
	\caption{Simulation results for learned BF decoding. In (a),
	results for standard BF (Alg.~\ref{alg:bit_flipping}) applied to $\Hoc$ overlap with
	hard-decision ML and are omitted. (BF: bit-flipping, WBF: weighted
	BF, LBF: learned BF (table Q-learning), LBF-NN: LBF with neural
	networks (fitted/deep Q-learning), s+d: sort and discard the
	channel reliabilities, HD ML: hard-decision maximum-likelihood, OSD: ordered statistics decoding)}
	\label{fig:results}
	\vspace{-0.3cm}
\end{figure*}

\begin{figure*}[t]
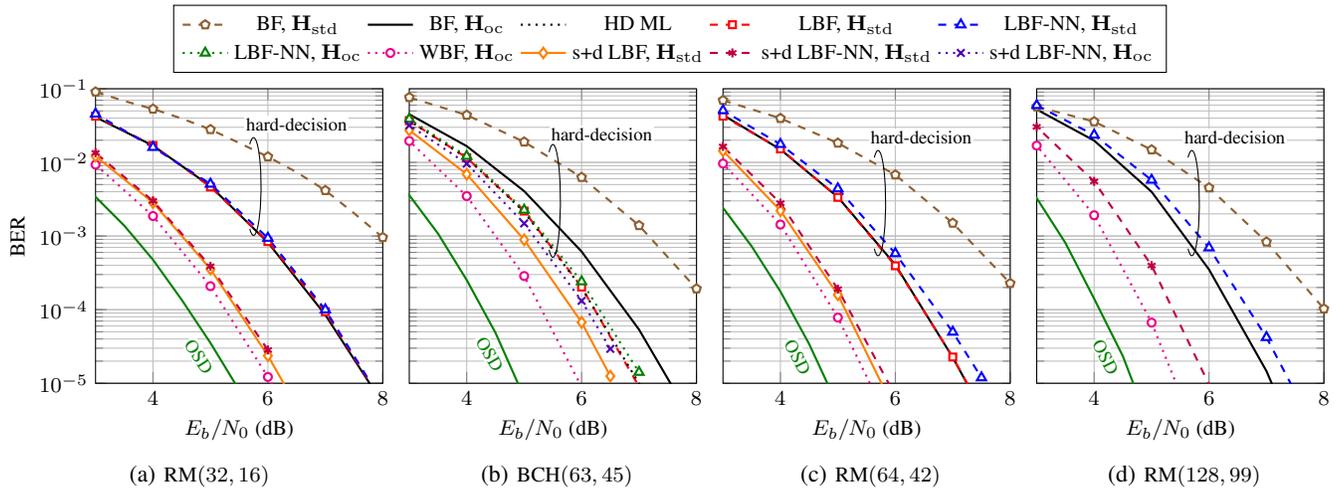

	\centering%
	\includestandalone{legend_CER}\\
	\subcaptionbox{\footnotesize RM$(32, 16)$}{\includestandalone{RM_2_5}}
	\subcaptionbox{\footnotesize BCH$(63, 45)$}{\includestandalone{BCH_63_45}}
	\subcaptionbox{\footnotesize RM$(64, 42)$}{\includestandalone{RM_3_6}}
	\subcaptionbox{\footnotesize RM$(128, 99)$}{\includestandalone{RM_4_7}}
	\caption{Bit error rate (BER) results for the same
	scenarios as considered in Fig.~\ref{fig:results}. }
	\label{fig:ber_results}
	\vspace{-0.3cm}
\end{figure*}

Fig.~\ref{fig:results} shows the CER performance for all
considered scenarios as a function of $\EbNo$. We start by focusing on
the ``hard-decision'' decoding cases, which are equivalent to assuming
transmission over the BSC. 
Supplementary bit error rate (BER) results for the same scenarios are shown in Fig.~\ref{fig:ber_results}.

\subsubsection{Baseline Algorithms} As a baseline for the LBF decoders
over the BSC, we use BF decoding according to
Alg.~\ref{alg:bit_flipping} (see also \cite[Alg.~II]{Bossert1986} and
\cite[Alg.~10.2]{Ryan2009}) applied to both the standard and
overcomplete PC matrices $\Hstd$ and $\Hoc$, respectively.  We also
implemented optimal syndrome decoding for RM$(32,16)$ and
BCH$(63,45)$. In general, BF decoding shows relatively poor
performance when applied to $\Hstd$, whereas the performance increases
drastically for $\Hoc$ (see also \cite{Bossert1986, Santi2018}). In
fact, for RM$(32,16)$, standard BF for $\Hoc$ gives virtually the same
performance as optimal decoding and the latter performance curves are
omitted from the figure. This performance increase comes at a
significant increase in complexity, e.g., for RM$(32,16)$, the
overcomplete PC matrix has $620$ rows compared to the standard PC
matrix with only $16$ rows. For the BCH code, there still exists a
visible performance gap between optimal decoding and BF decoding based
on $\Hoc$. 

\subsubsection{Q-learning} From Figs.~\ref{fig:results}(a) and (b), it
can be seen that the LBF decoders based on table Q-learning for
RM$(32,16)$ and BCH$(63,45)$ converge essentially to the optimal
performance. For RM$(64,42)$ in Fig.~\ref{fig:results}(c), the
performance of LBF decoding is virtually the same as for standard BF
decoding using $\Hoc$, which leads us to believe that both schemes are
optimal in this case. These results show that the proposed RL approach
is able to learn close-to-optimal flipping patterns given the received
syndromes. Note that for RM$(128,99)$, Q-learning would require a
table with $|\mathcal{S}| |\mathcal{A}| \approx 7\cdot 10^{10}$
entries which is not feasible to implement on our system.

\subsubsection{Fitted Q-learning} The main disadvantage of the
standard Q-learning approach is the large storage requirements of the
Q-table. Indeed, the requirements are comparable to optimal syndrome
decoding and this approach is therefore only feasible for short or
very-high-rate codes. Therefore, we also investigate to what extend
the Q-tables can be approximated with NNs and fitted Q-learning. The
number of neurons in the hidden layer of the NNs is chosen to be
$1500$ for RM$(128,99)$ and $500$ for all other cases.  The achieved
performance is shown in Fig.~\ref{fig:results}, labeled as ``LBF-NN''. For
the RM codes, it was found that good performance can be obtained using
fitted Q-learning using the standard PC matrix $\Hstd$. The
performance loss compared to table Q-learning is almost negligible for
RM$(32,16)$ and increases slightly for the longer RM codes. For the
BCH code, we found that fitted Q-learning works better using $\Hoc$
compared to $\Hstd$. For this case, the gap compared to optimal
decoding is less than $0.1\,$dB at a CER of $10^{-3}$.  

\subsection{AWGN Channel}

Next, we consider the AWGN channel assuming that the reliability
information is exploited for decoding. 

\subsubsection{Baseline Algorithms} Ordered statistics decoding (OSD)
is used as a benchmark, whose performance is close to ML
\cite{Fossorier1995}. In this paper, we use order-$\ell$ processing
where $\ell = 3$ in all cases. Furthermore, we employ WBF decoding
according to \cite[Alg.~10.3]{Ryan2009} using $\Hoc$. Similar to BF
decoding over the BSC, the performance of WBF is significantly better
for overcomplete PC matrices compared to the standard ones (results
for WBF on $\Hstd$ are omitted). From Fig.~\ref{fig:results}, WBF
decoding is within $0.6$--$1.1\,$dB of OSD for the considered codes.
We remark that there also exist a number of improved WBF algorithms
which may reduce this gap at the expense of additional decoding
complexity and the necessity to tune various weight and threshold
parameters, see \cite{Bossert1986, Kou2001, Zhang2004, Jiang2005,
Liu2005, Shan2005}. For RM codes of moderate length,
ML performance can also be approached using other techniques
\cite{Dumer2006}. 


\subsubsection{Q-Learning} As explained in Sec.~\ref{sec:sorted}, our
approach to LBF decoding over the AWGN channel in this paper consists
of permuting the bit positions based on $\vect{r}$ and subsequently
discarding the reliability values. For the RM codes, the particular
permutation strategy is described in Sec.~\ref{sec:sorted}. The
performance results for standard Q-learning shown in
Figs.~\ref{fig:results}(a) and (c) (denoted as ``s+d LBF'')
demonstrate that this strategy performs quite close to WBF decoding
and closes a significant fraction of the gap to OSD, even though
reliability information is only used to select the permutation and not
for the actual decoding. For the BCH code, we use the same permutation
strategy as described in \cite{Bennatan2018}. In this case, however,
the performance improvements due to applying the permutations are
relatively limited. 

\subsubsection{Fitted Q-Learning} For the NN-based approximations of
the Q-tables for the sort-and-discard approach, we use the NN
sizes from the previous section for the BSC. In this case,
fitted Q-learning obtains performance close to the standard
Q-learning approach for RM codes. Similar to the BSC, the performance
gap is almost negligible for RM$(32,16)$ and increases for the longer
RM codes. For RM$(128,99)$, sort-and-discard LBF decoding with NNs
closes roughly half the gap between soft-decision ML (approximated via
OSD) and hard-decision ML (approximated via BF on $\Hoc$).

\section{Conclusion}
\label{sec:conclusion}

In this paper, we have proposed a novel RL framework for BF decoding
of binary linear codes. It was shown how BF decoding can be mapped to
a Markov decision process by properly choosing the state and action
spaces, whereas the reward function can be based on a reformulation of
the ML decoding problem. In principle, this allows for data-driven
learning of optimal BF decision strategies. Both standard
(table-based) and fitted Q-learning with NN function approximators
were then used to learn good decision strategies from data.  Our
results show that the learned BF decoders can offer a range of
performance--complexity trade-offs.



\ifExternalBib

\bibliographystyle{IEEEtran}
\balance
\bibliography{WCLabrv,WCLnewbib,$HOME/Dropbox/lib/bibtex/library_mendeley}

\else

\fi

\end{document}